\documentclass{pasj00}
\draft
\begin{document}
\SetRunningHead{H. Nakajima et~al.}{
  Performance of the charge injection capability of Suzaku XIS}
\Received{2006//}
\Accepted{2006//}
\title{Performance of the Charge Injection Capability of Suzaku XIS}
\author{Hiroshi \textsc{Nakajima},\altaffilmark{1}
Hiroya \textsc{Yamaguchi},\altaffilmark{1}
Hironori \textsc{Matsumoto},\altaffilmark{1} \\
Takeshi Go \textsc{Tsuru},\altaffilmark{1}
Katsuji \textsc{Koyama},\altaffilmark{1}\\
Hiroshi \textsc{Tsunemi},\altaffilmark{2}
Kiyoshi \textsc{Hayashida},\altaffilmark{2}
Ken'ichi \textsc{Torii},\altaffilmark{2}
Masaaki \textsc{Namiki},\altaffilmark{2}\\
Satoru \textsc{Katsuda},\altaffilmark{2}
Masayuki \textsc{Shoji},\altaffilmark{2}
Daisuke \textsc{Matsuura},\altaffilmark{2}
Tomofumi \textsc{Miyauchi},\altaffilmark{2}\\
Tadayasu \textsc{Dotani},\altaffilmark{3}
Masanobu \textsc{Ozaki},\altaffilmark{3}
Hiroshi \textsc{Murakami},\altaffilmark{3} \\
Mark W. \textsc{Bautz},\altaffilmark{4}
Steve E. \textsc{Kissel},\altaffilmark{4}
Beverly \textsc{LaMarr},\altaffilmark{4}
Gregory Y. \textsc{Prigozhin}\altaffilmark{4}}
\altaffiltext{1}{%
   Department of Physics, Graduate School of Science, Kyoto University,
   Sakyo-ku, Kyoto, 606-8502}
\altaffiltext{2}{%
   Department of Earth and Space Science, Graduate School of Science,
   Osaka University, \\
   Machikaneyama-cho, Toyonaka, Osaka 560-0043}
\altaffiltext{3}{%
   Institute of Space and Aeronautical Science, 3-1-1, Yoshinodai,
   Sagamihara, Kanagawa, 229-8510}
\altaffiltext{4}{%
  Center for Space Research, Massachusetts Institute of Technology, \\
  77 Massachusetts Avenue Cambridge, MA 02139-4307, USA}
\maketitle
\begin{abstract}

A charge injection technique is applied to the X-ray CCD camera, XIS
(X-ray Imaging Spectrometer) onboard Suzaku.
The charge transfer inefficiency (CTI) in each CCD column (vertical transfer
channel) is measured by the injection of charge packets into a transfer
channel and subsequent readout.
This paper reports the performances of the charge injection capability
based on the ground experiments using a radiation damaged device, and
in-orbit measurements of the XIS.
The ground experiments show that charges are stably injected with the 
dispersion of 91~eV in FWHM in a specific column for the charges equivalent
to the X-ray energy of 5.1~keV. This dispersion width is significantly
smaller than that of the X-ray events of 113~eV (FWHM) at approximately
the same energy. The amount of charge loss during transfer in a specific
column, which is measured with the charge injection capability,
is consistent with that measured with the calibration source.
These results indicate that the charge injection technique can accurately
measure column-dependent charge losses rather than the calibration sources.
The column-to-column CTI correction to the calibration source spectra
significantly reduces the line widths compared to those with a column-averaged
CTI correction (from 193~eV to 173~eV in FWHM on an average at the time
of one year after the launch).
In addition, this method significantly reduces the low energy tail in the
line profile of the calibration source spectrum.

\end{abstract}

%
\section{Introduction}
\label{sect:intro}

The high positional and moderate energy resolutions of the Charged Couple 
Device (CCD) established this device to be the main detector for imaging
spectroscopy in X-ray astronomy since ASCA/SIS \citep{Burke93}.
However, one drawback to an X-ray CCD in-orbit is the degradation
of the gain and energy resolution due to an increase of the charge transfer
inefficiency (CTI).
The proton irradiation on the CCD chip increases the number of charge traps
in the CCD, which is composed of silicon crystals.
This defect is more severe for low energy protons because they deposit more
energy than high energy protons in the CCD transfer channel.
The main origin of the CTI and consequent gain degradation
is the increase of charge traps.
In fact, Chandra/ACIS has suffered from a degraded energy
resolution due to the low-energy ($\sim$~10 - 100~keV) protons in
the van Allen belts \citep{Plucinsky00}.
Although a thick shielding around the CCD camera can significantly
reduce the proton flux on the CCD, the radiation damage
cannot be ignored over a mission lifetime of several years.

In order to maintain the good performance of CCDs in orbit, the CTI
must be frequently measured and applied to the data.
Most of the major X-ray missions are provided with one or more
calibration sources to measure the CTI.
The number of charge traps is not uniformly distributed over the CCD
imaging area, and hence the CTI is also not uniform over the imaging area.
Therefore the CTI correction should be independently executed for each column
(vertical transfer channel).
However, the limited flux of calibration X-rays impedes an accurate
and frequent measurement of the CTI and its spatial variation over the imaging
area.

Recently a charge injection technique has been developed
\citep{Prigozhin04, Bautz04, LaMarr04, Smith04, Meidinger00}.
A charge packet with the amount of $Q$ is artificially injected through
a charge injection gate \citep{Tompsett75} into each column
and is subsequently readout as $Q'$ after the charge
transfer in the same manner as the X-ray event. This method allows us
to measure a charge loss ($\delta Q$ = $Q$ -- $Q'$) for each column,
which in turn, can potentially be a powerful tool for the CTI calibration. 

The X-ray Imaging Spectrometer (XIS; \cite{Koyama07} and references therein)
onboard the Japanese 5$^{\rm th}$ X-ray satellite Suzaku \citep{Mitsuda07} is
equipped with a charge injection structure
\citep{Prigozhin04, Bautz04, LaMarr04}.
The low earth orbit makes the detector background of the XIS lower and more
stable than those of Chandra and XMM-Newton. 
However, the XIS's gain and energy resolution have gradually degraded due to
the increase of the CTI during transit through the South Atlantic Anomaly.
After six-month from the first-light of the XIS, the CTI has increased to
non-negligible level. This result has stimulated us to investigate the
in-orbit charge injection performances. This paper reports on the results.

Section \ref{sect:about_ccd} and \ref{sect:about_CI} of this paper describe
the XIS and charge injection capability.
Section \ref{sect:experiments} is devoted to
the CTI experiments, while section \ref{sect:cal_results} describes the results
of the ground and onboard experiments. The discussion and summary are
in section \ref{sect:discussion} and \ref{sect:summary}, respectively.
The mean ionization energy of an electron by an X-ray in silicon 
is assumed to be $3.65~{\rm eV\ e^{-1}}$ throughout this paper. 

%
\section{The CCDs of the XIS}
\label{sect:about_ccd}

\citet{Koyama07} have provided details on the XIS and CCDs
(MIT Lincoln Laboratory model CCID41). Hence, we briefly duplicate
for the charge injection study of this paper. The CCDs are the three-phase
frame transfer type and have basically the same structure as those of
Chandra/ACIS. Each pixel size is 24~$\mu$m $\times$ 24~$\mu$m and the
number of the pixels is 1024 $\times$ 1024 in the imaging area.
Therefore, the size of the imaging area is $\sim$ 25~mm $\times$ 25~mm.
The exposure time is 8~sec for the normal clocking mode.
With the radiative cooling and a Peltier cooler, the CCD temperature is
controlled to --90~$^{\circ}$C. Hence, the dark current is suppressed
to $\sim$2 electrons/8sec/pixel.
Four CCDs are onboard Suzaku. Three of them are the
front-illuminated (FI) chips, while the other is the back-illuminated (BI)
chip. The BI chip has the same basic specifications as the FI chips,
except that the BI chip has a larger quantum efficiency in the soft energy
band. The ground calibrations verified that the thickness
of the depletion layer is
$\sim$~65~$\mu$m for the FI chips, and $\sim$~42~$\mu$m for the BI chip.
In order to see the function of the CCD, we give the schematic view of the
XIS FI chip in figure~\ref{fig:ccd_pic_sche}. Each CCD chip has four segments
(from A to D), and each segment has one readout node.
$^{55}$Fe calibration sources, which irradiate the upper edge of the
segment A and D, are used for the monitoring of the gain, CTI, and
energy resolution in orbit.

%
\section{The Charge Injection}
\label{sect:about_CI}

\begin{figure}
  \begin{center}
    \hspace{-1.0cm}
    \FigureFile(70mm,43.75mm){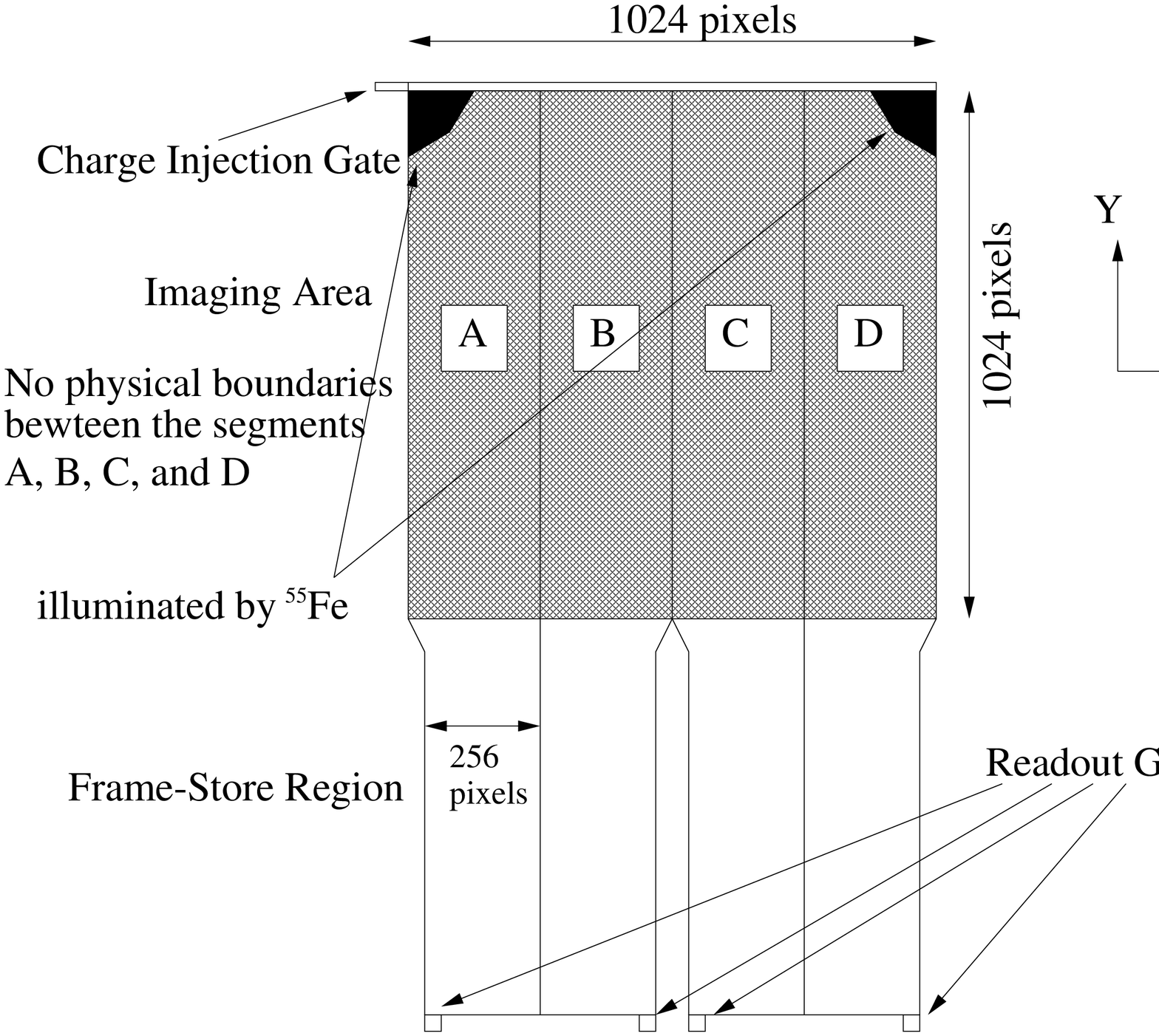}
  \end{center}
  \caption {Schematic view of the XIS FI chip.
    For the charge injection, a serial register is equipped at the top of
    the imaging area. The chip has four readout nodes, one for each segment.
    Signals are simultaneously read from these nodes.
    For the BI chip, the charge injection register runs from segment D to A.
    This figure is adopted from \citet{Koyama07}.
  }
  \label{fig:ccd_pic_sche} 
\end{figure}

\begin{figure*}
  \begin{tabular}{cc}
    \begin{minipage}{0.6\textwidth}
      \hspace{0.5cm}
      \FigureFile(100mm,40mm){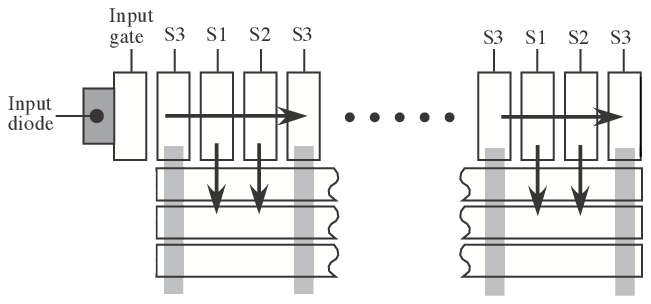}
    \end{minipage}
    \begin{minipage}{0.4\textwidth}
      \hspace{0.5cm}
      \FigureFile(40mm,50mm){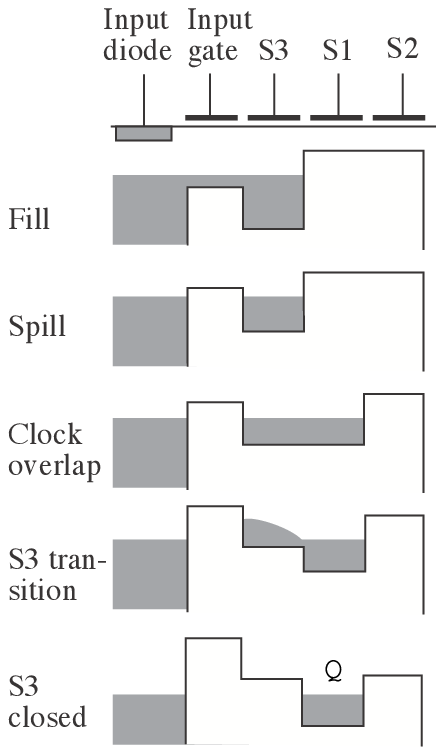}
    \end{minipage}
  \end{tabular}
  \caption{(Left panel:) Schematic view of the charge injection structure.
    Charge packets are injected to
    the charge injection register from an input gate located at the edge
    of the register. After depositing the packets over the register,
    a vertical clock runs to inject the packets into the imaging area.
    (Right panel:) Schematic view for the injection of the charges
    at the input gate.
    The offset voltage between the input gate and S3 electrode
    can control the amount of injected charge ($Q$).
    These figures are adopted from \citet{Bautz04}.
  }
  \label{fig:ci_schematic} 
\end{figure*}

\citet{Prigozhin04} have reported details of the charge injection structure.
By referring to figures~\ref{fig:ccd_pic_sche} and \ref{fig:ci_schematic},
here we describe the essential function of the charge injection.
For the brevity to describe the charge injection technique and its results,
notations of parameters, which will be frequently used in this paper are listed
in table~\ref{tab:abbr}.

\begin{table*}[h]
  \begin{center}       
    \caption{The notation list of parameters.}
    \label{tab:abbr}
    \begin{tabular}{ll}
      \hline
      \hline
      Parameters & Notation  \\ 
      \hline
      Injected charge (for one column) & $Q$ ($Q_{\rm{COL}}$) \\ 
      Readout charge (for one column) & $Q'$ ($Q'_{\rm{COL}}$) \\ 
      Charge loss in the transfer (for one column) &
      $\delta Q$ ($\delta Q_{\rm{COL}}$) \\
      Charge Transfer Inefficiency (for one column) & CTI (CTI$_{\rm{COL}}$)\\
      column-dependent CTI obtained with charge injection & CTI$_{\rm{CI}}$ \\
      column-averaged CTI obtained with the cal. source & CTI$_{\rm{CAL}}$ \\
      \hline
    \end{tabular}
  \end{center}
\end{table*}

\begin{figure*}
  \begin{tabular}{cc}
    \begin{minipage}{0.5\textwidth}
      \hspace{-1.5cm}
      \FigureFile(120mm,80mm){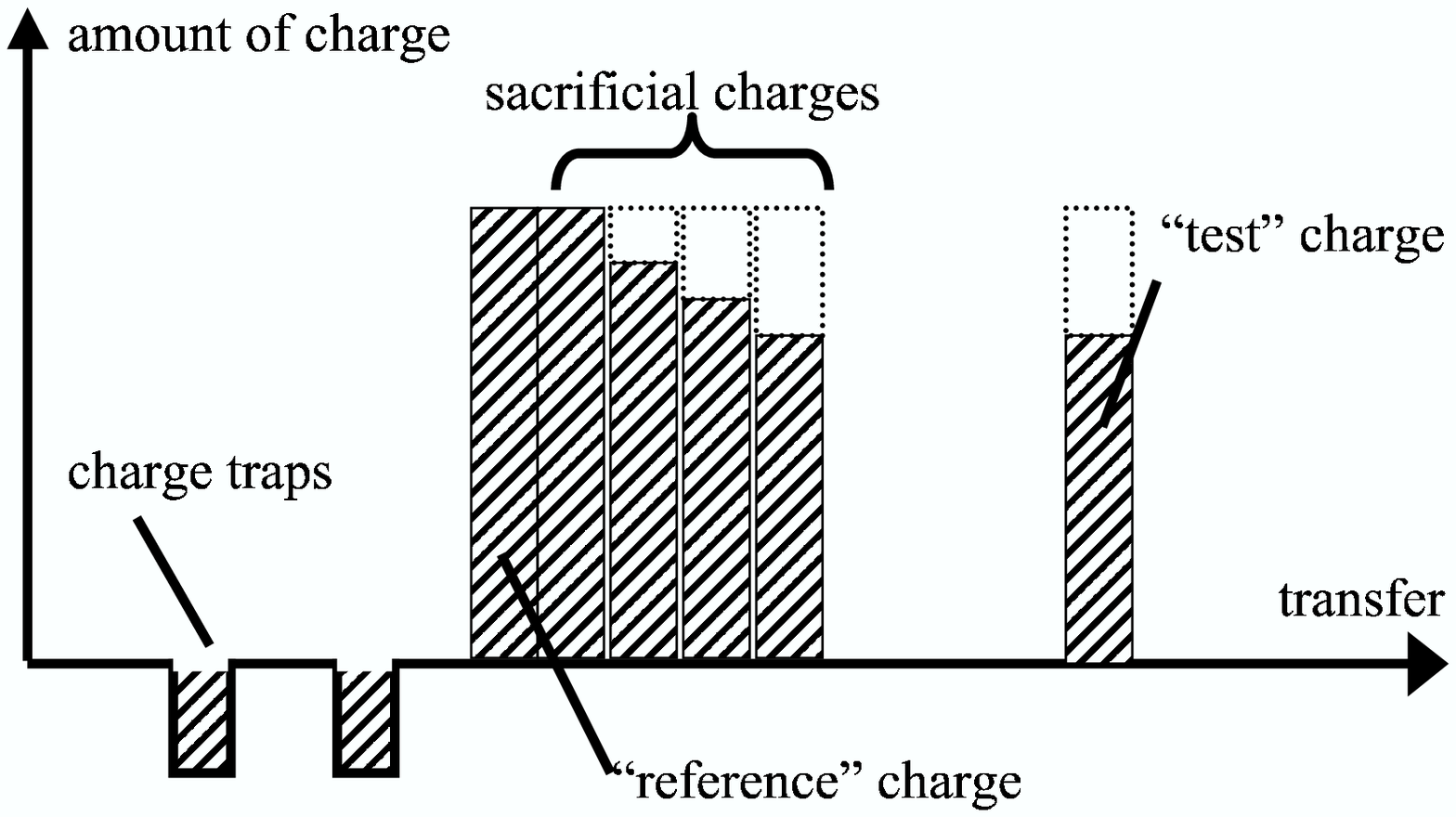}
    \end{minipage}
    \begin{minipage}{0.5\textwidth}
      \hspace{-1.0cm}
      \FigureFile(90mm,65mm){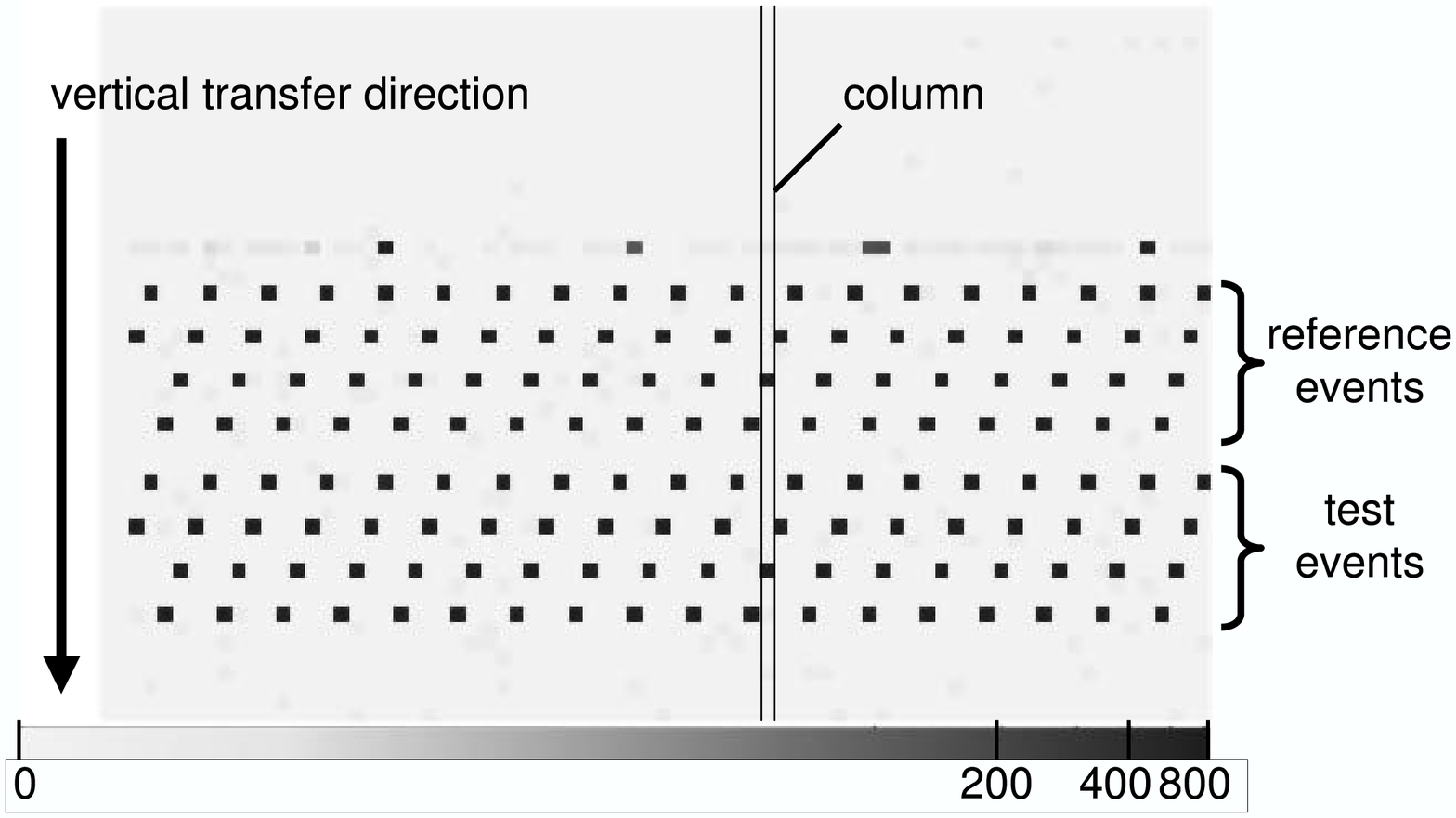}
    \end{minipage}
  \end{tabular}
  \vspace{-1.0cm}
  \caption{(Left panel:) Schematic view of how to measure $\delta Q$ using the
    {\it test} and {\it reference} charge packets.
    The amount of lost charge ($\delta Q$) can be estimated by comparing
    the PHA of {\it test} packet, which suffer from CTI, to that of
    {\it reference} packet, which is not affected by the traps.
    (Right panel:) Frame image of the XIS during the charge injection
    experiment. The gray scale shows the pixel level, not the number of events.
    There are two events per each column: The lower one is the {\it test}
    event and the upper one is the {\it reference} event.
    Note that the sacrificial packets are not displayed at the
    request of the event detection algorithm.
  }
  \label{fig:ciimage} 
\end{figure*}

A serial register of 1024 pixels long is attached to the next of the upper
edge of the imaging area (hereafter we call this charge injection register).
An input gate is equipped at left of the charge injection register
(see figure~\ref{fig:ccd_pic_sche}).
Pulling down the potential for electrons at the input gate and the next
electrode (S3 in figure~\ref{fig:ci_schematic}), the potential well
is filled with charges with the amount of $Q$. Then pulling up the potential,
the charge packet is spilled. The amount of charge is controlled by
the offset voltage between the input gate and the next electrode (S3).
In the normal XIS operations, this fill-and-spill cycle is repeated every
1/40960~sec $\simeq$ 24~$\mu$sec.
The deposited charge packets ($Q$s) in the charge injection register are
vertically transferred into the imaging area by the same clocking pattern
as that of X-ray events. A part of the packet ($\delta Q$) will be trapped
by the charge traps during the transfer.
After the launch, because $\delta Q$ is not negligible due to the increase
of charge traps, only $Q'$ can be measured with a injection of single charge
packet and the normal readout. On the other hand, we need the measurement
of $\delta Q$ in order to estimate the CTI.
We hence adopt the following injection pattern, with which we can obtain
both values of $Q'$ and $Q$ simultaneously
as shown in the left panel of figure~\ref{fig:ciimage}.
After injecting a {\it test} charge packet of $Q$ in one row (horizontal
transfer channel), we inject packets with the same amount of $Q$ in five
subsequent rows: the preceding four packets
are called the {\it sacrificial} charge packets and the last one is the
{\it reference} charge packet. The {\it test} charge packet is separated
from trains of {\it sacrificial} charge packets to allow the event detection
algorithm \citep{Koyama07}.

The {\it test} charge packet may suffer from traps in the transfer channel
(column), and therefore, the readout charge ($Q'_{test}$) should be
$Q$ -- $\delta Q$.
On the other hand, since the preceding {\it sacrificial} charge packets may
fill the charge traps, the subsequent {\it reference} charge packet may
not be trapped if the 
clocking time is shorter than the de-trapping time scale \citep{Gendreau93}.
Thus, the readout charge ($Q'_{ref}$) from the {\it reference} charge packet
should be approximately equal to $Q$.
The right panel of figure~\ref{fig:ciimage} shows a frame image taken
during our experiments.
The positions of the charge packet trains are periodically shifted
by one column to allow the proper event detection algorithm.
The {\it sacrificial} charge packets are not read because of the same reason.
The value $\delta Q$ after the transfer can be measured by subtracting the
mean pulse height amplitude (PHA) of the {\it test} events from that of the
{\it reference} events.
By selecting different $Q$s, we can also investigate the relation between
$Q$ and $\delta Q$.

%
\section{Ground and In-Orbit Experiments} 
\label{sect:experiments}

\subsection{Ground Experiments} 
\label{subsec:ground_cal_results}

\begin{figure}
   \begin{center}
     \FigureFile(80mm,50mm){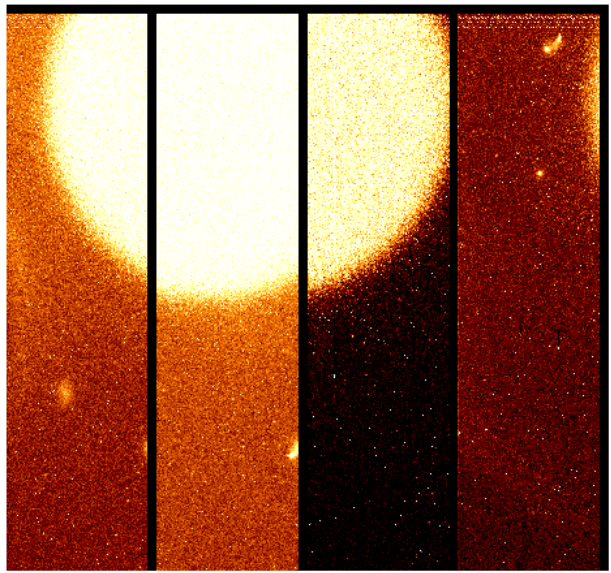}
   \end{center}
  \caption{Frame image of the proton-damaged FI chip used in the ground
    experiments. The gray scale is the pixel level, not the number of events.
    Gaps between segments are horizontal over-clock regions.
    The chip is irradiated by protons over the circular region that overlaps
    segments A, B and C. The pixel levels in the damaged region
    are systematically different from those in other regions
    due to the dark current.
  }
  \label{fig:frameimg} 
\end{figure}

Before the launch of Suzaku, we conducted the ground experiments with the same
type of CCD chip as the XIS in order to verify the performance of charge
injection function.
The CCD chip was damaged by protons utilizing the cyclotron at the
Northeast Proton Therapy Center at Boston (USA).
Proton beam of 40~MeV was irradiated on the circular region
shown in figure~\ref{fig:frameimg}. The total fluence 
was 2.0 $\times$ 10$^9$~cm$^{-2}$, which is approximately the same as that
the XIS may receive during several years in orbit. 
Experiments with damaged and non-damaged chips were conducted
in MIT and Kyoto University respectively, using the fluorescent X-ray
generation system (\citet{Hamaguchi00} for the latter)
and a $^{55}$Fe radioisotope.
During these experiments, the sensors were maintained at
a pressure of $\sim$10$^{-6}$~Torr and a CCD temperature of --90~$^{\circ}$C.
In this paper, we report on the results of the  FI chip,
because the quantitative differences between the FI and BI chips are small.

\subsection{In-Orbit Experiments} 
\label{subsec:onboard_cal_results}

The in-orbit charge injection experiment was conducted in the observations of
the supernova remnant (SNR) 1E0102--72.3 in the Large Magellanic Cloud.
All the data were acquired with the normal clocking mode and with 
the 3$\times$3 or 5$\times$5 editing modes \citep{Koyama07}.
We applied several values of $Q$ in order to investigate
the dependance of $\delta Q$ on $Q$.
Table~\ref{tab:explog} summarizes the experimental logs.

\begin{table*}[h]
  \begin{center}       
    \caption{The log of the charge injection experiment in orbit.} 
    \label{tab:explog}
    \begin{tabular}{lcc}
      \hline\hline
      \rule[-1ex]{0pt}{3.5ex}       & XIS0 \& XIS2      & XIS1 \& XIS3  \\
      \hline
      \rule[-1ex]{0pt}{3.5ex}  Date & 2006/7/17         & 2006/6/26-27    \\
      \rule[-1ex]{0pt}{3.5ex}  Time & 06:06:50 - 21:39:46 &
      02:47:07 - 02:37:55   \\
      \hline
      \rule[-1ex]{0pt}{3.5ex}  The equivalent X-ray energy &
      0.6/4.2/8.0 for XIS0 & 0.3/7.3 for XIS1 \\
      \rule[-1ex]{0pt}{3.5ex}  of the injected charge packets  (keV) &
      0.6/3.9/7.8 for XIS2 & 0.5/4.6 for XIS3 \\
      \hline
      \rule[-1ex]{0pt}{3.5ex}  Total Effective Exposure (ksec) & 6.0 & 5.2 \\
      \hline
      \multicolumn{3}{@{}l@{}}{\hbox to 0pt{\parbox{85mm}{\footnotesize
	    Notes. The XIS1 is the BI chip and the others are the FI chips.
	  }\hss}}
    \end{tabular}
  \end{center}
\end{table*}

%
\section{Measuring the Charge Loss for Each Column}
\label{sect:cal_results}

\subsection{Stability of the Amount of Injected Charge}
\label{subsec:stability}

\begin{figure}
  \begin{center}
    \FigureFile(60mm,42.86mm){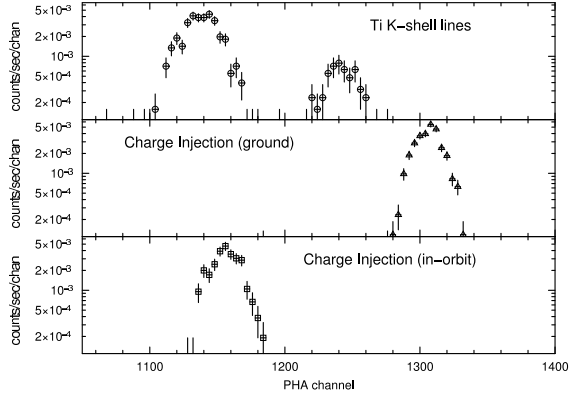}
  \end{center}
  \caption{(Upper panel:) Spectra of Ti K$\alpha$ fluorescent X-rays
    from an exposure of 350 frames taken in the ground experiments.
    (Middle panel:) Spectra of charge injection events collected on ground from
    an exposure of 800 frames. (Bottom panel:) Same as middle panel but
    collected in orbit from an exposure of 600 frames. The PHA level is
    slightly different from that in middle panel because the offset voltage
    is varied from that adopted in ground experiments.
    For all panels, the events are extracted from column of X=2. 
    Horizontal axis is the dark-level-subtracted PHA.
  }
  \label{fig:stability}
\end{figure}

In order to reliably estimate $\delta Q$, $Q$ in {\it reference} and
{\it test} events must be equal, because $\delta Q$ is the difference
between $Q'_{ref}$ and $Q'_{test}$.
If $Q$ can be controlled more accurately than that of charge dispersion
of X-ray events in each column, the charge injection offers obvious advantages
over conventional CTI measurements using X-ray calibration sources.
We hence check the stability of $Q'_{test}$ when a designed offset voltage
is applied at the input gate. For this propose, we use the non-damaged chip,
because $Q'_{test}$ should be nearly equal to $Q$ due to the negligible 
number of charge traps.
Figure~\ref{fig:stability} shows the spectra of fluorescent X-rays
(Ti K$\alpha$) collected on ground and $Q'_{test}$ of
approximately the same equivalent energy as X-rays collected both on ground
before proton damage and in orbit after the damage.
Events are extracted from one arbitrary column for all the data sets.
The FWHM of the $Q'_{test}$ is 91$\pm$4 eV for ground data and
95$\pm$6 eV for in-orbit data, that are significantly better than
that of the X-ray data of 113$\pm$7 eV.  The former FWHMs mean the stability
of $Q_{\rm{COL}}$ of the charge injection under the controlled offset voltage
at the input gate, and the latter is primarily due to the Fano noise.
Thus, we verify that $Q_{\rm{COL}}$ is sufficiently stable to estimate 
$\delta Q_{\rm{COL}}$.

\begin{figure*}
  \begin{tabular}{cc}
    \begin{minipage}{0.5\textwidth}
      \FigureFile(80mm,60mm){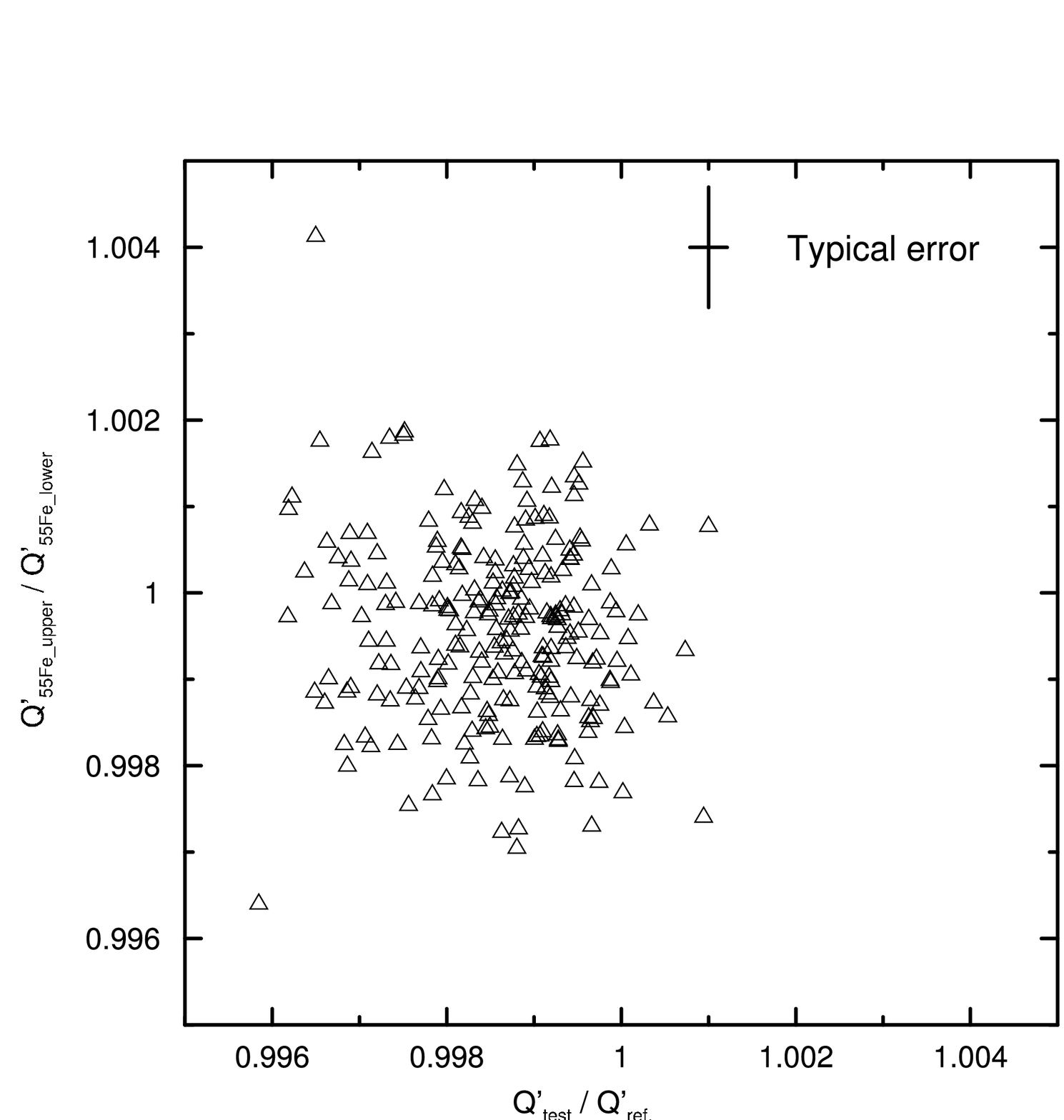}
    \end{minipage}
    \begin{minipage}{0.5\textwidth}
      \FigureFile(80mm,60mm){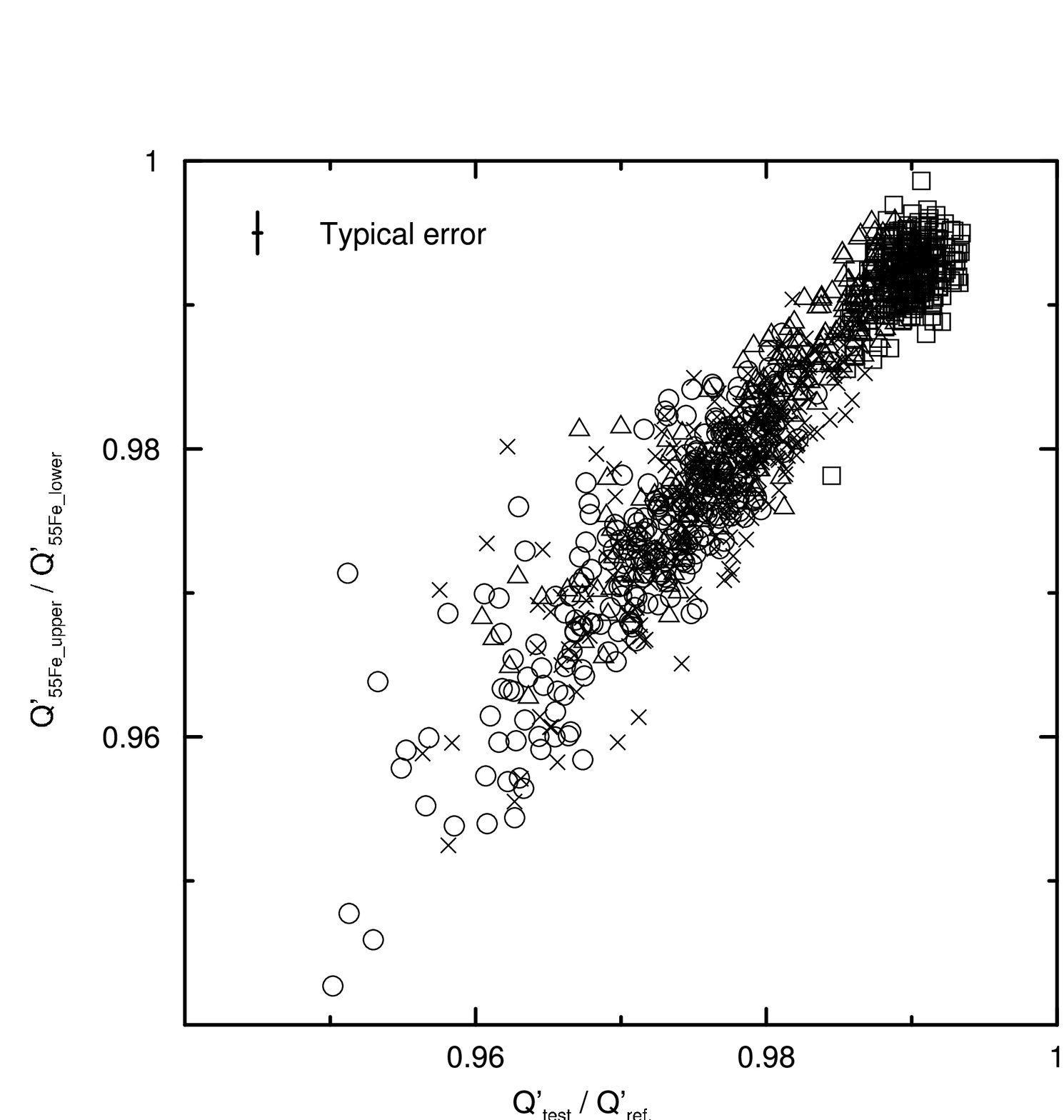}
    \end{minipage}
  \end{tabular}
  \caption{Ratio $Q'_{test}/Q'_{ref}$ (horizontal axis) and that of
    $Q'_{\rm{^{55}Fe\_upper}}$ and $Q'_{\rm{^{55}Fe\_lower}}$ (vertical axis)
    for each column.
    Open triangles, circles, crosses, and squares represent the columns in
    segment A, B, C, and D, respectively.
    (Left panel:) Plots for non-damaged chip. Note that plots for segments
    B, C, and D have the same distribution as segment A.
    Hence, segments B, C, and D
    are ignored from the figure for simplicity.
    (Right panel:) Plots for a proton-damaged chip.
    For both panels, anomalous columns including a hot or flickering pixel
    are eliminated. For all panels, typical errors indicate one sigma
    confidence levels.
  }
  \label{fig:correlation} 
\end{figure*}

Next, we investigate whether the ratio of $Q'_{test}/Q'_{ref}$
is proportional to the CTI, which is measured with the
$^{55}$Fe calibration X-rays. The PHA histograms of $Q'_{test}$
and $Q'_{ref}$ are fitted with a single Gaussian for each column.
For the $^{55}$Fe events, we extract the events from upper and lower 100
rows of the imaging area ($Q'_{\rm{^{55}Fe\_upper}}$
and $Q'_{\rm{^{55}Fe\_lower}}$). The 100 rows are
selected for the statistical point of view in the spectral fitting.
Figure~\ref{fig:correlation} shows
the correlations between $Q'_{test}/Q'_{ref}$ and
$Q'_{\rm{^{55}Fe\_upper}}/Q'_{\rm{^{55}Fe\_lower}}$.
For the non-damaged chip (the left panel), because $\delta Q$
during the parallel transfer is 0 or 1 electron, the correlation can be
hardly seen. For the damaged-chip, on the other hand, the increase of CTI is
significant in the circular region as shown in figure~\ref{fig:frameimg}.
We can see a clear positive correlation between the $^{55}$Fe and charge
injection events, especially in segment A, B, and C.
The best-fit slope is $\sim$1.05 and the correlation coefficient is 0.94
(d.o.f.=976). Hence, $Q'_{test}/Q'_{ref}$
properly reflects the CTI$_{\rm{COL}}$.

\subsection{Measuring and Compensating the Charge Loss}

Based on the verification for the charge injection technique in section
\ref{subsec:stability}, we apply this technique to the onboard data.

\begin{figure*}
  \begin{tabular}{cc}
    \begin{minipage}{0.5\textwidth}
      \FigureFile(60mm,50mm){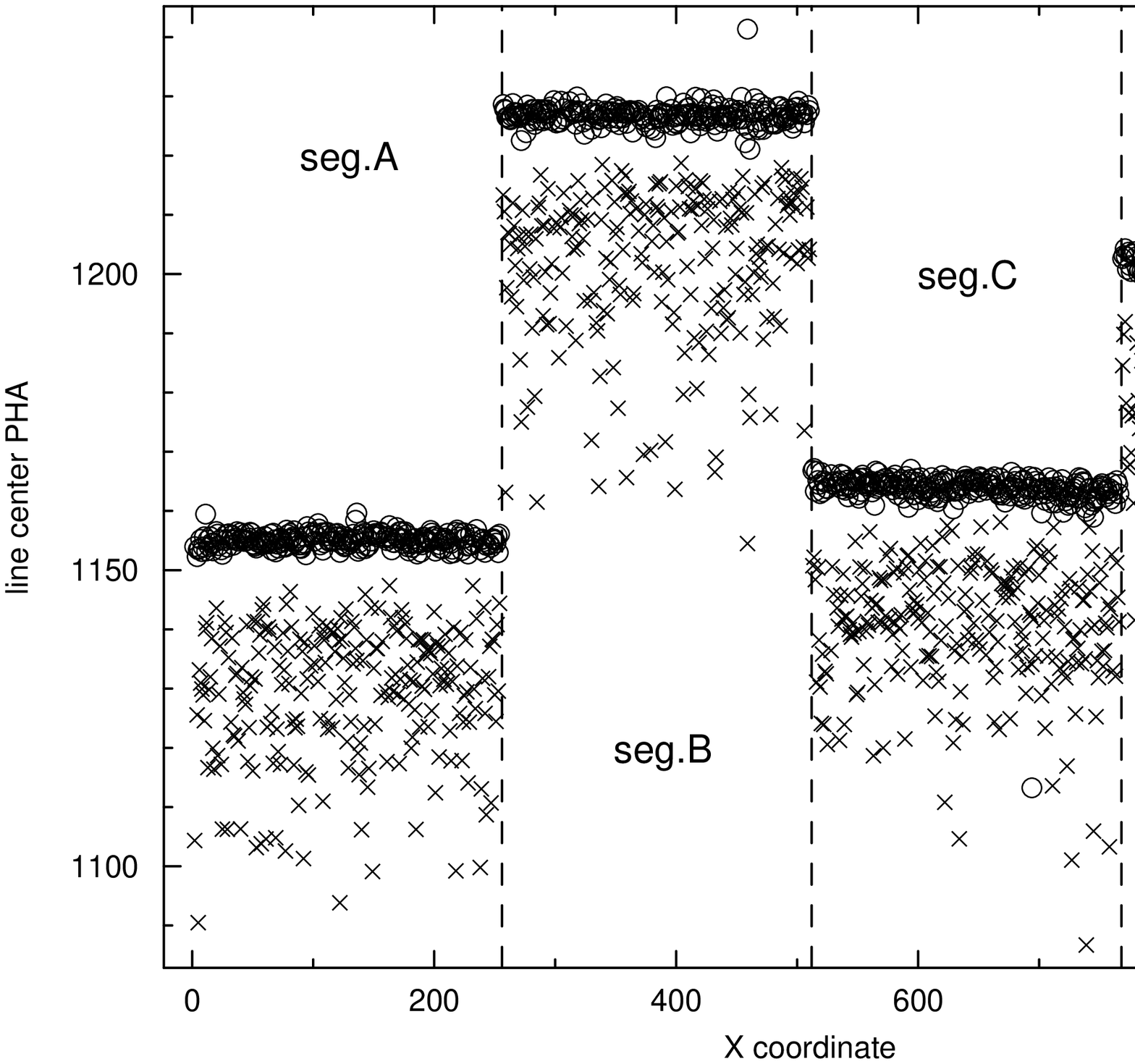}
    \end{minipage}
    \begin{minipage}{0.5\textwidth}
      \FigureFile(60mm,50mm){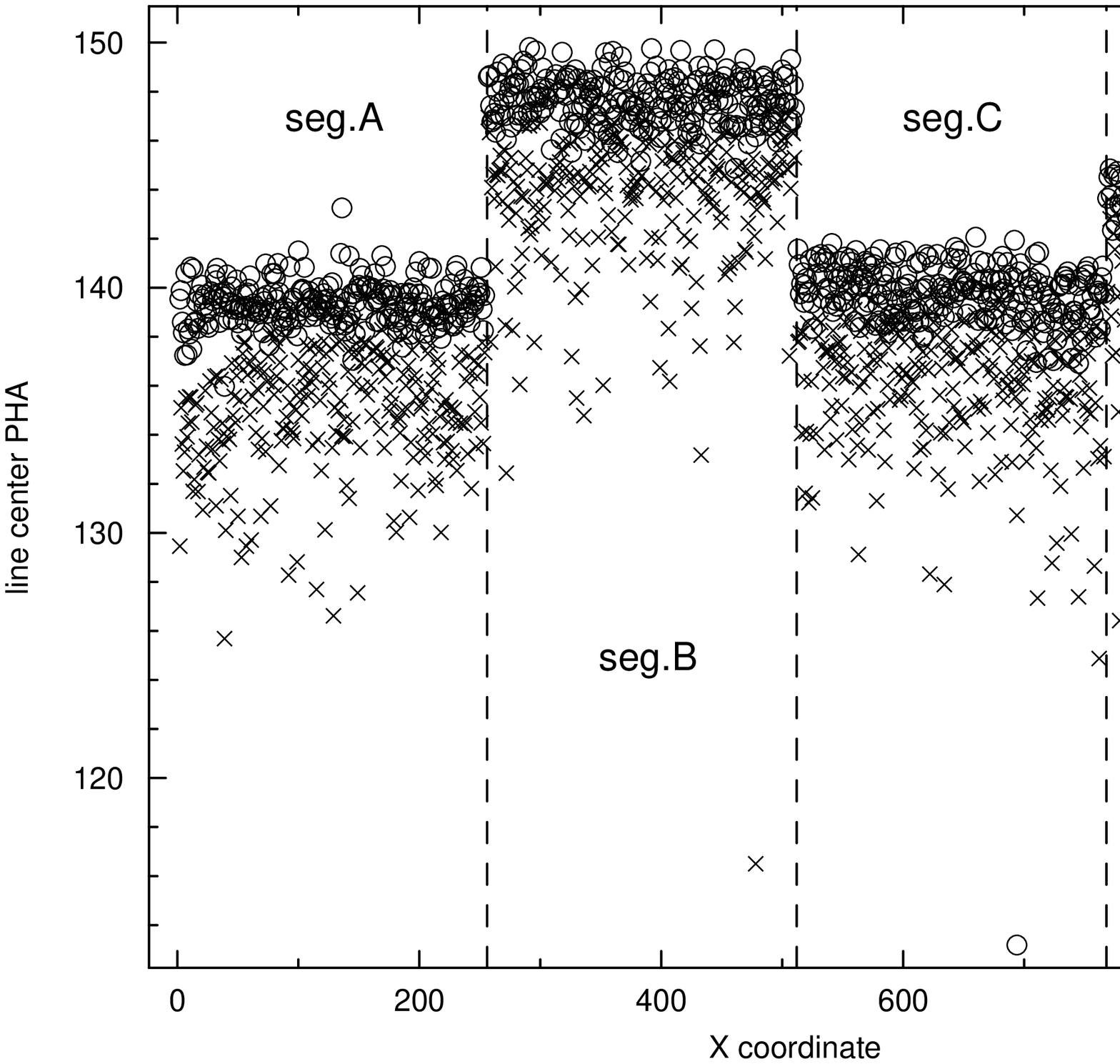}
    \end{minipage}
  \end{tabular}
  \caption{PHA distribution of the {\it test} (crosses) and {\it reference}
    (open circles) events as a function of the X coordinate of the XIS3 for two
    different $Q$s. The equivalent energies of these $Q$s are shown in
    table~\ref{tab:explog}.
    Note that because each segment has its own readout transistor and
    analog-to-digital converter, the gain varies from segment to
    segment, and hence, the PHA level varies.
    Anomalous columns including a hot or flickering pixel are eliminated.
  }
  \label{fig:reftest} 
\end{figure*}

\begin{figure}
  \begin{center}
    \FigureFile(70mm,50mm){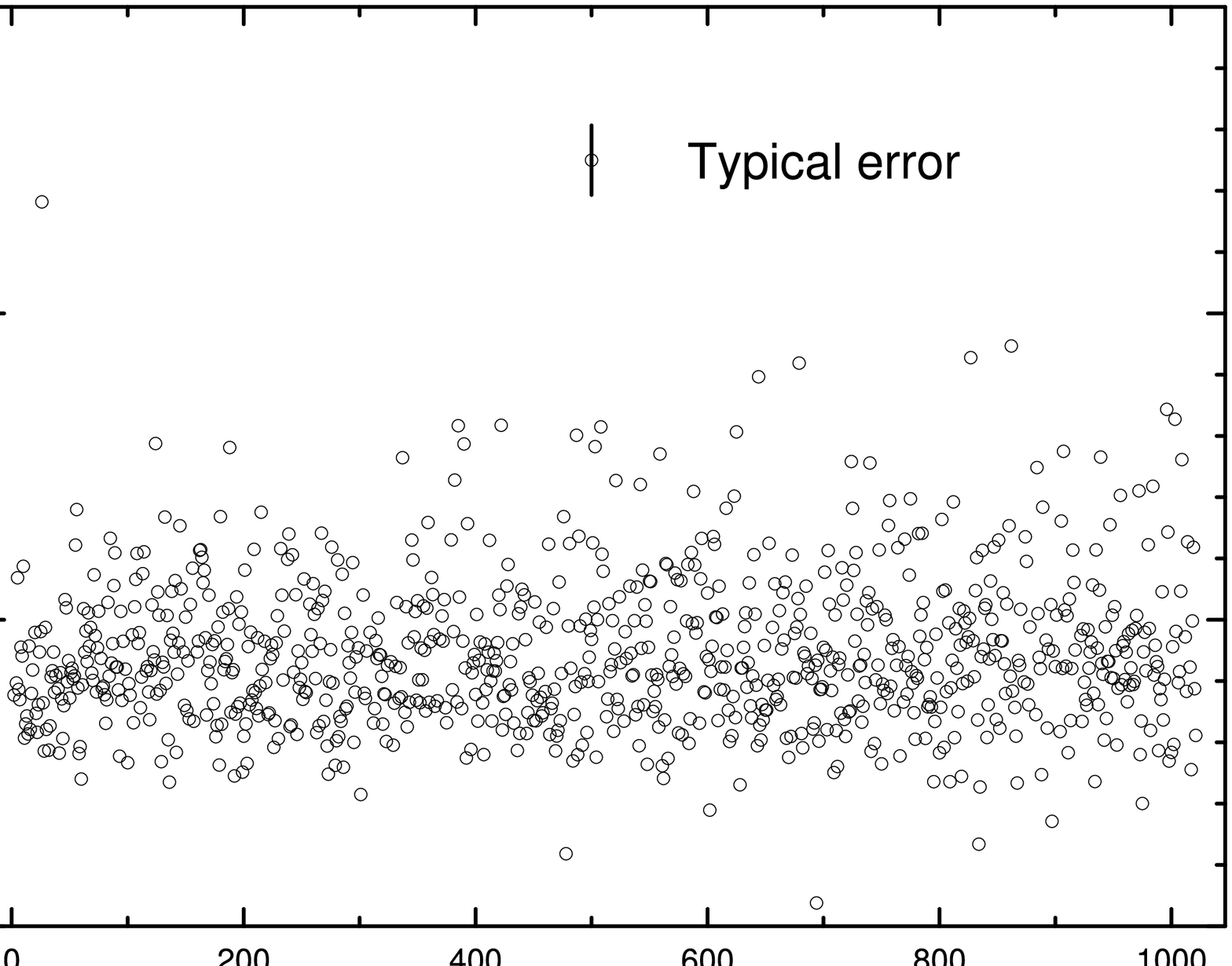}
  \end{center}
  \caption{Distribution of the power-index ($\alpha$) in the relation
    of $\delta Q \propto Q^{\alpha}$ (XIS3) as a function of the X
    coordinate, which is obtained through the charge injection experiments
    with the various charge amounts given in table~\ref{tab:explog}. Some
    anomalous columns are eliminated. Typical error indicate one sigma
    confidence level.
  }
  \label{fig:lostvspha} 
\end{figure}

Figure~\ref{fig:reftest} shows the PHA distribution for $Q'_{ref}$
(open circle) and $Q'_{test}$ (cross) as a function of X-coordinate (column).
The $\delta Q_{\rm{COL}}$ is clearly observed
in orbit for the first time. In order to estimate the $\delta Q$ dependance
on $Q$, we selected two or three different $Q$ values for the in-orbit charge
injection experiment (table~\ref{tab:explog}).
Assuming the single power law function of 
$\delta Q \propto Q ^{\alpha}$ as in \citet{Grant04}, we derive
$\alpha$ for each column. The results are given in figure~\ref{fig:lostvspha}.
The weighted mean values of $\alpha$ are 0.62, 0.71, 0.62, and 1.00 for
the XIS0, 1, 2, and 3, respectively. These values are roughly
consistent with another ground experiment \citep{Prigozhin04}.

\begin{figure}
  \begin{center}
    \FigureFile(60mm,42.86mm){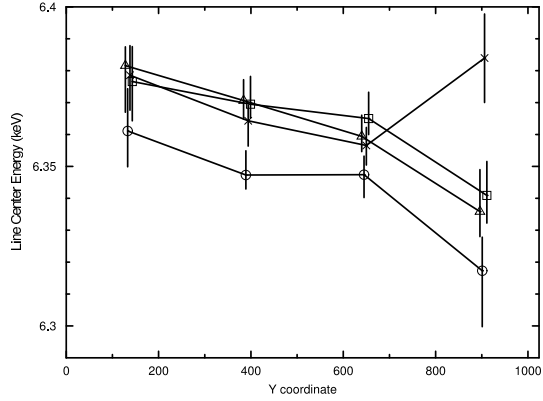}
  \end{center}
  \caption{The Y coordinate dependance of the center energy of 6.4~keV line
    emission from the Sgr~C region. Photons are each extracted from 1024
    $\times$ 256 region.
    Open triangles, circles, crosses, and squares represent the plots of the
    XIS0, 1, 2, and 3, respectively. The error bars indicate 90\% confidence
    levels.
  }
  \label{fig:sgrcactydep} 
\end{figure}

The charge injection data provide only the information on the $\delta Q$
at the edge of the imaging area and hence we need to know the
Y coordinate dependance of $\delta Q$ from the data of celestial objects
that extend over the field of view of the XIS.
Figure~\ref{fig:sgrcactydep} shows the center energy of the 6.4 keV line as
a function of the Y coordinate for the diffuse X-rays from the Sgr~C region 
(Obs. Sequence = 500018010, Obs. Date = 2006-02-20).
The line center at the lower edge of the image (Y=0) deviates from
6.40 keV due to the CTI
during the fast transfer of all the data in the imaging area to 
the frame-store region (hereafter we call this frame-store-transfer).
However, the origin of the deviations at the other image regions
is complicated because the charges suffer from three kinds of the CTI:
the CTI in the imaging area due to the frame-store-transfer, that in
the frame-store region due to the the frame-store-transfer and that
in the frame-store region due to the subsequent slow vertical transfer.
The CTI during the horizontal transfer is ignored in this work.
The CTI depends on the number density of charge traps and the transfer speed
\citep{Hardy98}. The shielding depth and pixel size are different in
the imaging area and frame-store region. The CTIs therefore may be different
between these areas. However, we cannot estimate each CTI component separately
from the total CTI seen in figure~\ref{fig:sgrcactydep}.

\begin{figure}
  \begin{center}
    \FigureFile(80mm,33.33mm){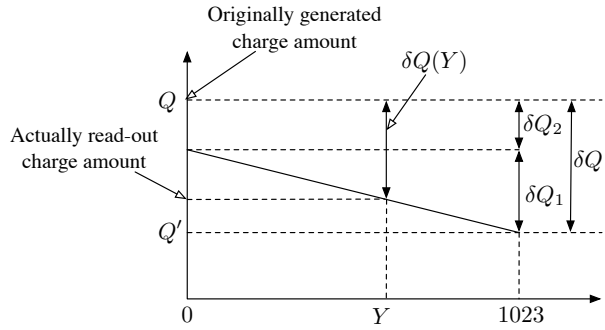}
  \end{center}
  \caption{The phenomenological relationship between the charge amount
    of $Q$ generated by an X-ray absorbed at $Y$ and the column-averaged
    charge loss of $\delta Q(Y)$.
    The amount of read-out charge is given by $Q-\delta Q(Y)$.
    $\delta Q_1$ and $\delta Q_2$ are the components dependent on and
    independent of  Y, respectively.}
  \label{fig:deltaQ} 
\end{figure}

Due to this limitation, we assume following phenomenological
compensation of the charge as shown in figure~\ref{fig:deltaQ}.
The charge loss during the transfer is assumed to consists of a component
depending on Y ($\delta Q_1$) and a Y-independent one
($\delta Q_2$). Both components are proportional to $Q^\alpha$, but
their proportionality constants may be different from each other.
Considering that the $Q$ is generated by an X-ray absorbed at $Y$,
the column-averaged charge loss of $\delta Q(Y)$ is given by
\begin{eqnarray*}
  &&\delta Q_1 = A_1\times Q^\alpha,
  \ \ \ \delta Q_2 = A_2\times Q^\alpha,\\
  &&\delta Q(Y)=\left[\delta Q_1 \cdot\frac{Y}{1023}+\delta Q_2\right].
\end{eqnarray*}
$A_1$ and $A_2$ can be estimated from figure \ref{fig:sgrcactydep}.
We next determined the column-dependent charge loss of $\delta Q_{\rm COL}(Y)$
so that the following relation holds at any Y coordinate.
\begin{eqnarray*}
  &&\delta Q_{\rm COL}(Y)=\delta Q(Y) \times
  \frac{\delta Q_{\rm COL}(1023)}{\delta Q(1023)},
\end{eqnarray*}
where $\delta Q_{\rm COL}(1023)$ can be estimated from
figure~\ref{fig:reftest}.
Hence we can compensate charge correctly in the entire region of
the imaging area.

\subsection{Energy Resolution}

\begin{figure}
  \begin{center}
    \FigureFile(60mm,42.86mm){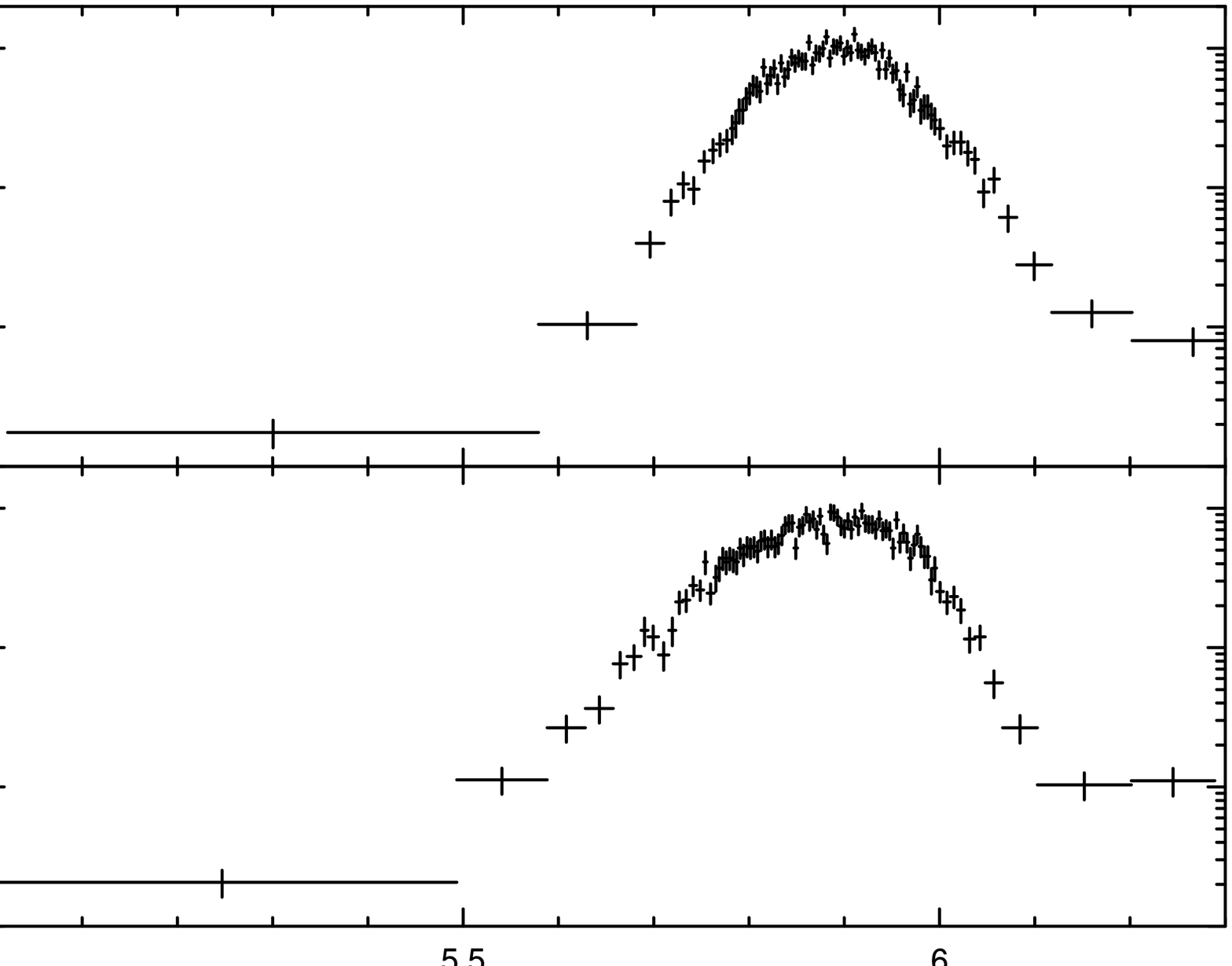}
  \end{center}
  \caption{Energy spectra of the calibration source of segment D of the XIS2
    after the CTI$_{\rm{CI}}$ correction (upper panel) and
    the CTI$_{\rm{CAL}}$ correction (lower panel).
    Data is simultaneously obtained with the charge injection experiments.
    The low energy tail component is significantly reduced by the charge
    injection technique.
  }
  \label{fig:cispectra} 
\end{figure}

Without a CTI correction, the energy resolution gradually decreases
at a rate of $\sim$~50~eV in FWHM @ 5.9~keV per year. Using
the $\delta Q_{\rm{COL}}$ determined with the charge injection experiment, 
we make the new spectra for the calibration sources in the observation 
given in table~\ref{tab:explog}.
Figure~\ref{fig:cispectra} shows the calibration source spectra after
the CTI$_{\rm{CI}}$ (upper) and CTI$_{\rm{CAL}}$ (lower) correction. 
The tail component after the CTI$_{\rm{CI}}$ correction is
significantly reduced compared to that after the CTI$_{\rm{CAL}}$
correction. This strongly indicates that the origin of the tail component
is the dispersion of the CTI$_{\rm{COL}}$ among columns. Hence
the temporal variation in the response function
can be suppressed by the charge injection technique.
Figure~\ref{fig:eneres} shows the FWHM of the calibration source
spectra after the CTI$_{\rm{CI}}$ and CTI$_{\rm{CAL}}$ correction.
On average, the FWHM is significantly improved from 193 eV to 173 eV.
These are the first in-orbit results for the charge injection function.

\begin{figure}
  \begin{center}
    \FigureFile(55mm,39.29mm){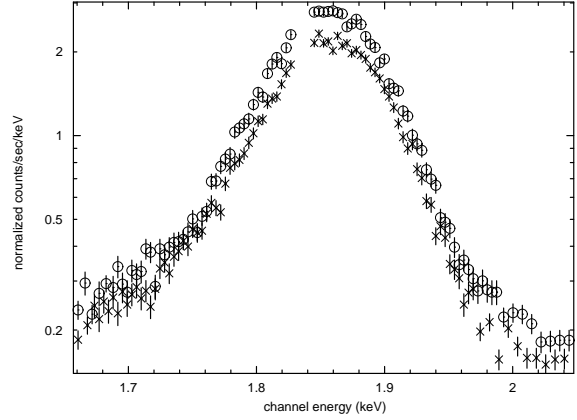}
  \end{center}
  \caption{Spectra around the He-like Si K$\alpha$ emission line of the
    west part of Tycho's SNR (XIS3). Data represented by crosses is
    charge-compensated with the CTI$_{\rm{CAL}}$, while the other
    (open circles) is compensated with the CTI$_{\rm{CI}}$.
    Note that the former is multiplied by 0.8 to avoid confusion.
    The data around the Si K-edge (1.839~keV) is ignored because
    the calibration at this energy has still large error.
  }
  \label{fig:tychosika}
\end{figure}

Our final purpose is to demonstrate that $\delta Q_{\rm{COL}}$
parameters are effective for celestial objects.
We apply the $\delta Q_{\rm{COL}}$ parameters to the Tycho's SNR
data (Obs. Sequence = 500024010, Obs. Date = 2006-06-27).
Figure~\ref{fig:tychosika} shows the spectra around the He-like Si K$\alpha$
emission line in the west part of Tycho's SNR for the XIS3
after the CTI$_{\rm{CI}}$ correction and CTI$_{\rm{CAL}}$
correction.
We see the same benefit as shown in figure~\ref{fig:cispectra}.
Note that the latter is multiplied by 0.8 to avoid confusion.
Radiation damage continuously increases while in orbit, and hence,
the benefit of the charge injection technique will become more apparent
over time as shown in this figure.

%
\section{Discussion}
\label{sect:discussion}

\begin{figure}
  \begin{center}
    \FigureFile(60mm,42.86mm){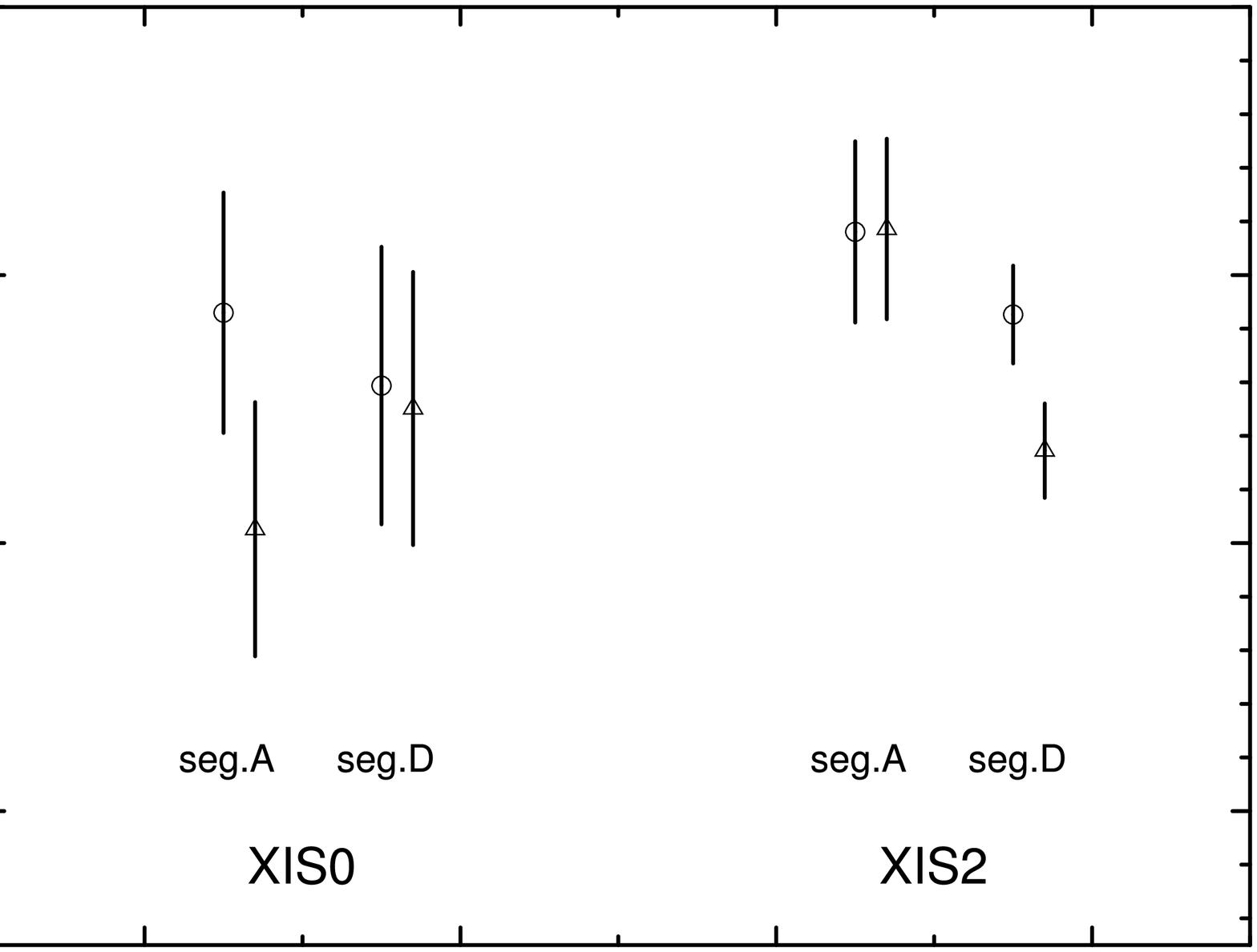}
  \end{center}
  \caption{FWHMs of the calibration source spectra simultaneously obtained with
    the charge injection experiment in May 2006. The $\delta Q_{\rm{COL}}$
    parameters obtained in May (open triangles) and July 2006 (open circles)
    are applied.
  }
  \label{fig:longterm}
\end{figure}

Because the CTI$_{\rm{CI}}$ correction parameters are time dependent,
$\delta Q_{\rm{COL}}$ must be periodically measured.
We make two sets of CTI$_{\rm{CI}}$ using $\delta Q_{\rm{COL}}$
derived from the charge injection experiment in May and July 2006,
and apply to the calibration source data taken in May 2006 for the XIS0
and XIS2. The results are shown in figure~\ref{fig:longterm}.
The average CTI in July is normalized to that in May,
while the relative variation in CTI$_{\rm{CI}}$ is preserved.
While the observation time of the charge injection and $^{55}$Fe data differs
by two months, the FWHM is significantly degraded only for the brightest
calibration source.
This confirms that, in practical sense, the interval of two months between
each CTI measurement is sufficient because there are few celestial
objects which has emission line brighter than this calibration source.

\begin{figure}
  \begin{center}
    \FigureFile(60mm,42.86mm){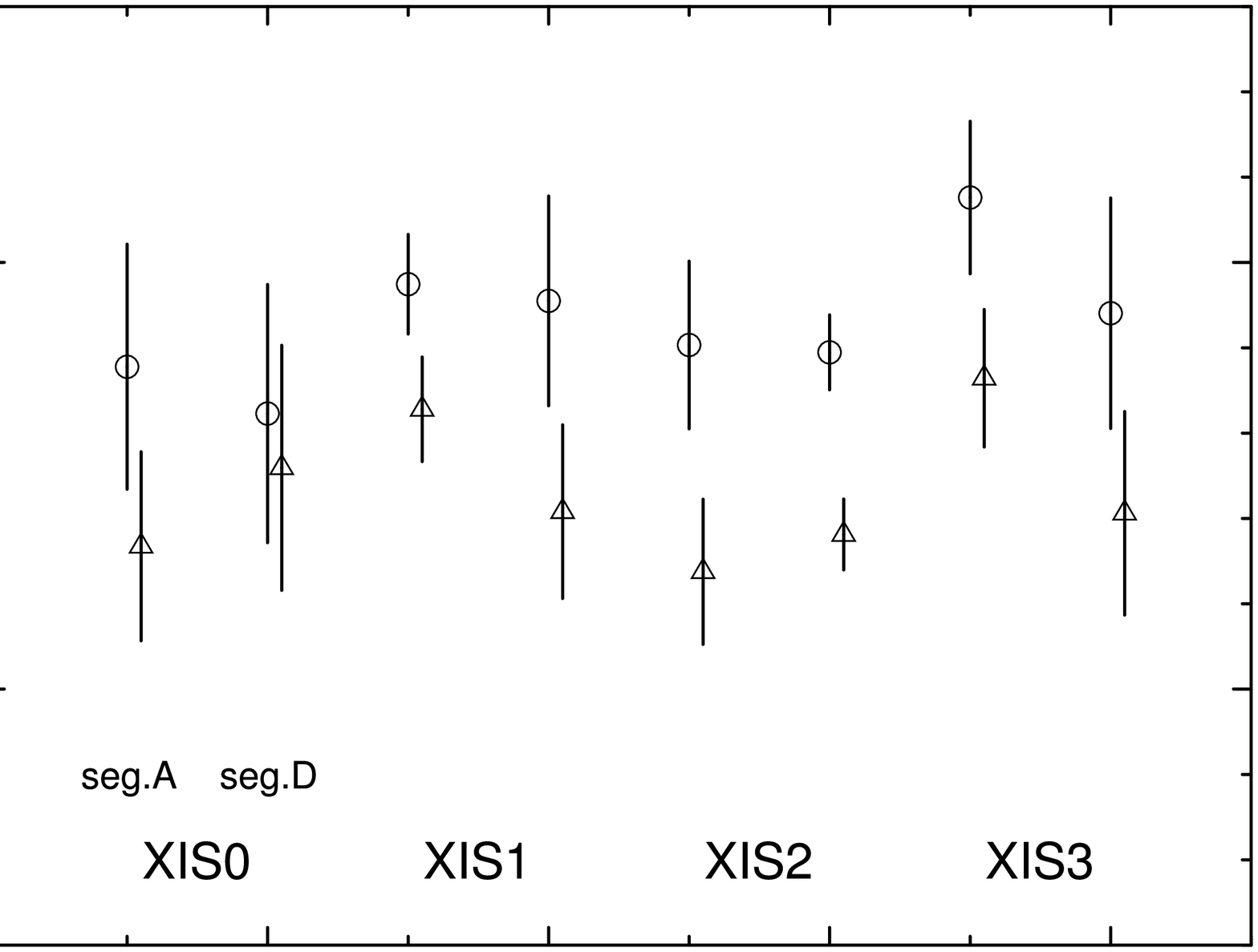}
  \end{center}
  \caption{FWHMs of the calibration source spectra simultaneously obtained
    with the charge injection experiments.
    Open triangles and circles represent the data after the correction
    with the CTI$_{\rm{CI}}$ and CTI$_{\rm{CAL}}$, respectively.
    The error bars indicate 90\% confidence level.
  }
  \label{fig:eneres}
\end{figure}

\begin{figure}
  \begin{center}
    \FigureFile(60mm,42.86mm){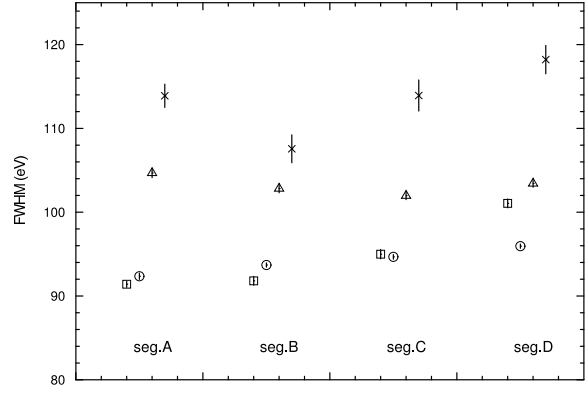}
  \end{center}
  \caption{FWHMs of the charge injection {\it test} events spectra before
    (cross) and after (open triangle) the correction of CTI$_{\rm{CI}}$,
    and those of the
    {\it reference} events (open circle) for each segment of the XIS3.
    FWHMs of the {\it test} events collected on ground are also shown
    (open square).
    Even after the CTI correction, the FWHMs of the {\it test} events spectra
    are larger than those of the {\it reference} events due to the
    probability process of the charge trapping.
  }
  \label{fig:sighikaku}
\end{figure}

Although the charge injection technique significantly improves the energy
resolution of the calibration source spectra, the FWHMs, as shown in
figure~\ref{fig:eneres}, are still larger than those
before the launch (130~eV).
Because the charge trapping is a probability process, the number of trapped
electrons ($\delta Q$) may also have a probability deviation, which would
increase the line width after the transfer. Hence the CTI correction with
charge injection technique cannot completely restore the line broadening.
We confirm this effect in figure~\ref{fig:sighikaku}, which
shows the FWHM of the charge injection {\it reference} events and those of
the {\it test} events before and after the charge compensation for the XIS3
in-orbit data. In addition, the {\it test} events measured on ground
(before the radiation damage) are shown.
The smaller FWHMs in all segments of the {\it reference} events are due to the
fact that the {\it reference} events may not lose charge
because the charge traps are already filled by the {\it sacrificial} events.
In fact, the FWHMs of the {\it reference} events are consistent with
those of the {\it test} events collected on ground although segment D show
anomalous trend.

Another characteristic of figure~\ref{fig:sighikaku} is that the FWHMs
increase along with X coordinate for the {\it reference} events and
{\it test} events collected on ground. This is due to the CTI in the charge
injection register. This influences the accuracy of the mean PHA of the
{\it reference} events and hence the accuracy of $\delta Q$.
However, the CTI in the charge injection register is rather lower than that
in the imaging area and frame-store region because the charge packets are
injected with an interval of 3 pixels and hence they can
work as {\it sacrificial} charges \citep{Gendreau95}.
In fact, as shown in figure~\ref{fig:eneres}, there is no significant
difference between segment A and D for the improvement of the FWHM of
the calibration source spectra.

These results lead us to use the charge injection capability to fill
the traps in the transfer channel by periodically injection of $Q$.
A ground experiment
shows that charges injected into every 54$^{\rm th}$ row improve the energy
resolution \citep{Bautz04}. We are trying to utilize this charge injection
technique for the onboard observations.
The results will be presented in a separate paper.

%
\section{Summary} 
\label{sect:summary}

The results of ground and in-orbit experiments of the
charge injection capability of the XIS are as follows:

\begin{enumerate}

\item The amount of injected charge ($Q$) is sufficiently stable
(dispersion of 91~eV in FWHM),
which should be compared to the X-ray energy resolution (FWHM of 113~eV) with
the same amount of charge.

\item The CTI depends on PHA of the charge. The charge loss can be explained
as $\delta Q \propto Q^{0.62-1.00}$

\item With the CTI$_{\rm{CI}}$ correction, the energy resolution
(FWHM) of $^{55}$Fe is improved compared to that with the CTI$_{\rm{CAL}}$
correction (from 193~eV to 173~eV on an average) at the time of one
year after the launch and the tail component in the line
profile is also significantly reduced.

\item The improved charge compensation method is applied to
the Tycho's SNR data, which results in the same benefit as the
calibration source data.

\item We confirm that the energy resolution can be largely improved
by filling the charge trap. Hence, we are currently trying another in-orbit
charge injection capability experiment to actively fill the charge traps
in the transfer channel by the charge injection technique.

\end{enumerate}

\section*{Acknowledgment}

The authors express their gratitude to all the members of the XIS team.
HN and HY are financially supported by the Japan Society for the
Promotion of Science.
This work is supported by a Grant-in-Aid for the 21st Century COE
``Center for Diversity and Universality in Physics'' from Ministry of
Education, Culture, Sports, Science and Technology,
and by a Grant-in-Aid for Scientific Research on Priority Areas in Japan
(Fiscal Year 2002-2006) ``New Development in Black Hole Astronomy''.

%

\clearpage

\end{document}